\documentclass[aps,twocolumn,pra,showpacs,floatfix,nofootinbib]{revtex4-1}
\usepackage{epsfig}
\usepackage{graphicx}
\usepackage{dcolumn}
\usepackage{amssymb,amsmath}
\usepackage{mathrsfs}

    \usepackage{color}
    \usepackage{soul}

    \newcommand{\ket}[1]{\big| \,{#1}\, \big> }
    
    \newcommand{\expect}[1]{\big< \,{#1} \, \big> }
    \newcommand{\matrixe}[3]{\big< \,{#1}\, \big| \,{#2}\, \big| \,{#3}\, \big> }
    \newcommand{\op}[1]{\widehat{#1}}

    \newcommand{\Ac}{^{227}\!{\rm Ac}}

\begin{document}
 \title{Enhanced nuclear Schiff moment in stable and metastable nuclei}
 
 \author{V.V. Flambaum  $^{1,2,3}$} \email{Email address: v.flambaum@unsw.edu.au}
 \author{H. Feldmeier $^4$}

 

\affiliation{$^1$ School of Physics, University of New South Wales,  Sydney 2052,  Australia}
\affiliation{$^2$Helmholtz Institute Mainz, Johannes Gutenberg-Universit\"at, Mainz 55099 , Germany}
\affiliation{$^3$The New Zealand Institute for Advanced Study, Massey University Auckland, 0632 Auckland, New Zealand}
\affiliation{$^4$ GSI,  Helmholtzzentrum fur Schwerionenforschung,  Planckstrasse 1, 64291 Darmstadt, Germany}
\date{\today}

\begin{abstract}
Nuclei with static intrinsic  octupole deformation  or a soft octupole vibrational mode lead to strongly enhanced collective nuclear Schiff moments. 
Interaction between electrons and these Schiff moments  produce enhanced time reversal (T) and parity (P) violating  electric dipole moments (EDM) in atoms and molecules.
Corresponding experiments may be used to test CP-violation theories predicting T,P-violating nuclear forces  and to search for axions.
Nuclear octupole deformations are predicted in many short lived isotopes. 
This paper investigates octupole deformations in stable and very long lifetime nuclei such as  $^{153}$Eu, $^{235}$U, $^{237}$Np and $^{227}$Ac, which can ease atomic experiments substantially. The estimates of the enhanced Schiff moments, atomic electric dipole moments and $T,P$-odd interaction constants in molecules containing these nuclei are presented.

 \end{abstract}

\maketitle

\section{Octupole deformation, soft octupole vibrations
and enhanced nuclear Schiff moments} 

Measurements  of  T,P and CP violating electric dipole moments (EDM) of elementary particles, nuclei, and atoms  provide crucial tests of unification theories and have already cornered many popular models of CP-violation including supersymmetry \cite{PR,ERK}. 
But the corresponding effects are very small, therefore, we are looking for mechanisms that enhance the effects
 - see e.g.  \cite{Khriplovich,KL,GF}. 

According to the Schiff theorem the nuclear EDM is completely screened  in neutral atoms  \cite{Schiff}.  
EDMs of diamagnetic atoms and molecules  are produced by interaction of electrons with the nuclear Schiff moment. 
The Schiff moment is a vector multipole of the nuclear charge distribution.
It appears in the third order of the multipole expansion of the  nuclear electrostatic potential with added electron screening term (together with  the electric octupole moment) \cite{Sandars,Hinds,SFK,FKS1985,FKS1986}.
As the operator for the Schiff moment violates  
parity its expectation value vanishes for nuclear many-body states 
which possess 
good parity and angular momentum quantum numbers. 
However, if there is a yet undiscovered small parity and time reversal breaking part in the nucleon interaction a finite Schiff moment appears and the corresponding electric field is felt by the electrons.

The distribution of the Schiff moment electric field inside the nucleus (this field vanishes outside the nucleus), the Hamiltonian describing the interaction of the Schiff moment electric field with relativistic atomic electrons,  and the  finite nuclear size corrections to the formula for the Schiff moment have been considered in Ref. \cite{FG}. 
This Schiff moment electric field polarizes the  atom and produces  an  atomic EDM directed along the nuclear spin. 

Refs. \cite{Sandars,Hinds} calculated  the Schiff moment due to a proton EDM.   
Refs. \cite{SFK,FKS1985,FKS1986} calculated (and named) the nuclear Schiff moment  produced by the P,T-odd nuclear forces. 
It was shown in \cite{SFK} that the contribution of the P,T-odd forces to the nuclear EDM and Schiff moment is  larger than the contribution of a nucleon EDM.   

Enhancement  of the nuclear Schiff moment may be due to  close nuclear levels of opposite parity  but the same angular momentum. 
They can be mixed by T,P-odd nuclear forces \cite{SFK}.
\footnote{Nuclear EDM and magnetic quadrupole produced by the T,P-odd nuclear forces are also enhanced  due to close levels  \cite{HH}. Collective enhancement of the magnetic quadrupole moments in deformed nuclei have been demonstrated in \cite{F1994}.}   
However, the largest enhancement ($\sim 10^2 - 10^3$ times) happens in nuclei with an intrinsic octupole deformation where both, the 
small energy difference of nuclear levels with opposite parity and the collective effect work together \cite{Auerbach,Spevak}. Atomic and molecular EDMs produced by the Schiff moment 
increase with the nuclear charge $Z$ and vary rapidly, faster than $Z^2$  \cite{SFK}. 
Therefore, heavy atoms, especially actinides are advantageous.
    
The Schiff moment is defined by the following expression \cite{SFK}: 
\begin{equation}\label{S}
{\bf S}=\expect{\op{\bf S}}=\frac{e}{10} [\expect{r^2 {\bf r}} - \frac{5}{3Z}\expect{r^2}\expect{\bf r}] \ , 
\end{equation}
where $\expect{r^n} = \int \rho({\bf r}) r^n d^3r$ are the moments of the nuclear charge density $\rho$ 
and ${\bf r}$ is measured from the nuclear center-of-mass position.        
The second term originates from the electron screening and contains  
the
nuclear mean squared charge radius  $\expect{r^2}/Z$ and nuclear EDM  $d=e\expect{\bf r}$, where $Z$ is the nuclear charge. 

EDM and Schiff moment are polar T-even vectors which must be directed along the nuclear spin $\bf{J}$ which is a T-odd pseudovector. Therefore, in the absence of T,P-violation EDM and Schiff moment vanish.

The strong interaction is  parity conserving, therefore
the eigenstates of the nuclear Hamiltonian $\op{H}$ are also eigenstates of parity $\op{\Pi}$. 
\begin{align}\label{eq:eigen}
	\op{H}\,\ket{J^\pi M,n} = &\ E^{J^\pi}_n \ket{J^\pi M,n}\\
	\op{\Pi}\,\ket{J^\pi M,n} = &\ \pi \ \ket{J^\pi M,n}      \nonumber
\end{align}
A weak parity and time reversal violating interaction  $\op{W}$ mixes opposite parity states,
\begin{equation}\label{eq:states}
	\ket{'+',0}=\ket{+,0}+ \sum_{n\neq 0} \ket{-;n}\frac{\matrixe{-,n}{\op{W}}{+,0}}{E^+_0 - E^-_n}\,, 
\end{equation}
and produces the Schiff moment,
\begin{align}\label{eq:odd}
{\bf S} &=  \matrixe{'+',0}{\bf \op{S}}{'+',0} =\\
 &\sum_{n\neq 0}
\frac{2Re\big\{\matrixe{-,n}{\op{W}}{+,0}\matrixe{+,0}{\bf \op{S}}{-,n}\big\}}{E^+_0-E^-_n}\nonumber\ .
\end{align}
Here we use the short hand notation $\ket{\pm,n}\equiv\ket{J^\pm,M=J,n}$  for spin aligned eigenstates of the nuclear Hamiltonian 
$\op{H}$ which have the same spin but opposite parity.
Analogue for $\ket{'-',0}$ if the ground state has negative parity.
\footnote{An ordinary T-conserving, P-violating weak  interaction $\op{W}^P$ 
does not generate a Schiff moment
since the product
$\matrixe{-,n}{\op{W}^P}{+,0}\matrixe{+,0}{\bf \op{S}}{-,n}$ in Eq.~\eqref{eq:odd} is purely imaginary while
it is real for T,P-violating $\op{W}$.}

Eq.~\eqref{eq:odd} indicates that for a large Schiff moment 
the task is to find nuclei where the energy denominator $|E^+_0 - E^-_n|$ is small and the matrix element of ${\op{S}}$ is large. 
A collective enhancement of $\matrixe{+,0}{\bf \op{S}}{-,n}$ is expected in nuclei with intrinsic octupole deformation or a soft octupole vibration mode.
In section ~\ref{sec:exp-octupole} experimental evidence for collective octupole rotations is discussed by identifying corresponding rotational spectra. 

In Refs.~\cite{Auerbach,Spevak} it is shown that for a nucleus with an intrinsic
octupole deformation $\beta_3$ and a quadrupole deformation $\beta_2$ in the  body-fixed frame the Schiff moment $S_{intr}$ is proportional to the octupole moment $O_{intr}$, i.e. it  has a collective nature:
\begin{equation}\label{Sintr}
 S_{intr} \approx \frac{3}{5 \sqrt{35}} O_{intr} \beta_2 \approx    \frac{3}{20 \pi  \sqrt{35}} e Z R^3 \beta_2 \beta_3 ,
\end{equation}
where $R$ is the nuclear radius.   

A  nucleus with an 
intrinsic octupole deformation and non-zero total spin $J$ in the ground state has 
doublets of close opposite parity rotational states $\ket{J^{\pm}}$ with the same angular momentum $J$ which may be schematically presented as 
\begin{equation}\label{OmegaDoublet}
| J^{\pm} >=\frac{1}{\sqrt{2}} (|\Omega> \pm |-\Omega>),
\end{equation}
where $\Omega$ is  the projection of $J$ on to the nuclear axis (a more specific  definition of these states is presented in Eq.~\eqref{eq:proj}) . 
The states of this doublet are mixed by a P,T-violating interaction $\op{W}$. 
The mixing coefficient is:
\begin{equation}\label{alpha}
 \alpha=\frac{\matrixe{J^-}{\op{W}}{J^+}}{E^+  -  E^-} . 
\end{equation}
For simplifying the argument only the opposite parity state with the largest contribution in the sum of Eq.~\eqref{eq:states} is considered.
Calculation of the matrix element $<J^-|\op{W}| J^+>$ in this case is reduced to the calculation of the expectation value $<\Omega |\op{W}| \Omega> = -<-\Omega | \op{W} | -\Omega>$. 
According to Ref.~\cite{Spevak} the T,P-violating matrix element is approximately equal to
  \begin{equation}\label{W}
   \matrixe{J^-}{\op{W}}{J^+} \approx \frac{\beta_3 \eta}{A^{1/3}}  \textrm{eV}.
  \end{equation}  
Here $\eta$ is the dimensionless strength constant of the nuclear T,P-violating potential $W$:
   \begin{equation}\label{eta}
 W= \frac{G}{\sqrt{2}} \frac{\eta}{2m} ({\bf \sigma \nabla}) \rho\, ,
   \end{equation}
where $G$ is the Fermi constant, $m$ is the  nucleon mass and $\rho$ is the nuclear number density.

In states with good parity  in Eq.~(\ref{OmegaDoublet}) probabilities of the nuclear axis along the nuclear spin (state $|\Omega>$)  and opposite to the nuclear spin  (state $|-\Omega>$) are both equal to 1/2 and average orientation of the nuclear axis is zero,   $\expect{ n_z }=0$.

The mixing of the doublet states makes these probabilities different, 
$<\Omega| n_z |\Omega>$  appears with the probability $(1+\alpha)^2/2$ and 
$<-\Omega| n_z |-\Omega>$  appears with the probability $(1-\alpha)^2/2$, 
i.e. this mixing polarizes the  nuclear axis ${\bf n}$ along the nuclear spin ${\bf J}$,  
\begin{equation}\label{n}
 \expect{ n_z } = 2 \alpha \frac{J_z}{J+1},
\end{equation}
and the intrinsic Schiff moment shows up in the laboratory frame \cite{Auerbach,Spevak}. In the ground state with the maximal projection of the angular momentum $J_z=J$ the Schiff moment is \cite{Auerbach,Spevak} 
\begin{equation}\label{Scol}
 S= 2 \alpha \frac{J}{J+1} S_{intr}. 
\end{equation}
Eqs. (\ref{Sintr}-\ref{Scol}) give an analytical estimate for the Schiff moment in nuclei with octupole deformation \cite{Spevak,Th} which agrees with the more accurate numerical calculations in Ref. \cite{Spevak}:
 \begin{equation}\label{San}
 S \approx 1. \cdot 10^{-4} \frac{J}{J+1} \beta_2 \beta_3^2 Z A^{2/3} \frac{\textrm{keV}}{E^-  -  E^+} e \,\eta \, \textrm{fm}^3,
 \end{equation}
For example, this estimate gives $S=280 \cdot 10^{-8} \, e \,\eta \, \textrm{fm}^3$ for $^{225}$Ra, very close to the result  of the numerical calculation in Ref.  \cite{Spevak} $S=300 \cdot 10^{-8} \, e \,\eta \, \textrm{fm}^3$. Eq. (\ref{San}) gives values of the Schiff moment in nuclei with intrinsic
octupole deformation which are 2-3 orders of magnitude larger than the Schiff moments in spherical nuclei  
\footnote{The values of the Schiff moments for the nuclei  with octupole deformation listed above vary  from 45 to 1000 $10^{-8} e \eta$ fm$^3$   \cite{Spevak,Th}. For spherical nuclei  $^{199}$Hg, $^{129}$Xe, $^{203}$Tl and  $^{205}$Tl, where the Schiff moment measurements have been performed,  the calculations \cite{SFK,FKS1985,FKS1986} give the Schiff moment $S \sim 1 \times 10^{-8} $ $e \eta$ fm$^3$.}. 

 The Schiff moment in Eq.(\ref{San}) is proportional to the squared octupole deformation parameter $\beta_3^2$ which is about $(0.1)^2$. According to Ref. \cite{Engel2000}, in nuclei with a soft octupole vibration mode the squared dynamical octupole deformation  $\expect{\beta_3^2 } \sim (0.1)^2$, i.e. it is the same as the static octupole deformation. This means that a similar enhancement of the Schiff moment may be due to the dynamical octupole effect \cite{Engel2000,FZ,soft2} in nuclei where 
 $\expect{\beta_3}=0$. 
Calculations performed in Ref. \cite{FZ} have demonstrated that it is sufficient to replace the static  
$\beta_3^2$ by the dynamical $\expect{\beta_3^2 }$ in Eq. (\ref{San}) to obtain the estimate for the collective Schiff moment in the nuclei with the soft octupole mode. Another important observation is  that  in an oscillator  $\expect{\beta_3^2 } \propto 1/ \omega$, where in our case the octupole oscillator frequency $\hbar \omega=  |E^-  -  E^+|$, so the Schiff moment 
$S \propto 1/ |E^-  -  E^+|^2$. Thus, for the effect of the soft octupole vibration mode, the 
closeness of levels is even more important than for the static deformation.

   In the papers \cite{Auerbach,Spevak}  the numerical calculations of the Schiff moments and estimates of atomic EDM produced by electrostatic interaction between electrons and these moments have been done  for  $^{223}$Ra, $^{225}$Ra, $^{223}$Rn, $^{221}$Fr, $^{223}$Fr, $^{225}$Ac and $^{229}$Pa
\footnote{  Accurate relativistic many-body  calculations of atomic  EDM induced by the Schiff moment in Hg, Xe, Rn, Ra and Pu atoms have been performed in Refs. \cite{SFK,FlambaumRa,DzubaRa,AtomicSchiff,AtomicSchiff2}.}. 
According to Refs. \cite{Auerbach,Spevak} the Schiff moment of  $^{225}$Ra exceeds the Schiff moment of $^{199}$Hg (where the most accurate  measurements of the Schiff moment have been performed \cite{HgEDM}) 200 times. 
Even larger enhancement of the  $^{225}$Ra Schiff moment has been obtained  in Refs.~\cite{EngelRa,Jacek2018}.  
      
Within the meson exchange theory the $\pi$-meson exchange gives the dominating contribution to the T,P-violating nuclear forces \cite{SFK}. According to Ref. \cite{FDK} the neutron and proton constants in the T,P-odd potential  (\ref{eta}) may be presented as  $\eta_n \approx - \eta_p \approx 5 \times 10^6 (  -0.2 g {\bar g}_0 +  g {\bar g}_1 +  0.4 g {\bar g}_2$), where $g$ is the strong $\pi$-meson - nucleon  interaction constant and ${\bar g}_0$, ${\bar g}_1$, ${\bar g}_2$ are the  $\pi$-meson - nucleon CP-violating interaction  constants in the isotopic channels $T=0,1,2$.
The most sophisticated  calculation of the $^{225}$Ra Schiff moment  has been done in Ref.  \cite{EngelRa}  where they presented the Schiff moment as 
$S( ^{225}\textrm{Ra})= (a_0 g {\bar g}_0 +  a_1 g {\bar g}_1 +  a_2 g {\bar g}_2) e $ fm$^3$. To estimate the error the authors  of Ref. \cite{EngelRa} have done the calculations using 4 different models of the strong interaction. They obtained the following 4 sets of the coefficients: $a_0= -1.5,\, -1.0,\, -4.7,\, -3.0$; 
   $a_1= 6.0,\, 7.0,\, 21.5,\, 16.9$;  $a_2= -4.0,\, -3.9,\, -11.0,\, -8.8$.    Taking the average values of the coefficients gives:
  \begin{equation}\label{SRag}
 S( ^{225}\textrm{Ra})= (  - 2.6 g {\bar g}_0 +  12.9 g {\bar g}_1 -6.9 g {\bar g}_2)\, e\, \textrm{fm}^3,\\
  \end{equation}   
The analytical formula in Eq. (\ref{San}) allows us to scale the value of the Schiff moment to other nuclei using the numerical calculation result Eq. (\ref{SRag})  for $^{225}$Ra which has  $\beta_3$=0.099, 
  $\beta_2$=0.129, $J=1/2$ and interval between the opposite parity levels $E(1/2^-)- E(1/2^+)$=55.2 keV. We present the result as  
  \begin{equation}\label{Sg}
S(g) = K_S (  - 2.6 g {\bar g}_0 +  12.9 g {\bar g}_1 -6.9 g {\bar g}_2)\, e\, \textrm{fm}^3,
  \end{equation}
 where $K_S=K_J K_{\beta}K_A K_E$, $K_J=\frac{3 J}{J+1}$, $K_{\beta}=791\beta_2 \beta_3^2$, $K_A=0.00031 Z A^{2/3}$,  $K_E= \frac{55\textrm{keV}}{E^-  -  E^+} $. By definition, all of these coefficients are equal to 1 for  $^{225}$Ra and are of the order  of unity for other heavy nuclei with octupole deformation.
 
 One can express the results in terms of more fundamental parameters such as the QCD $\theta$-term constant  ${\bar \theta}$ and the  quark chromo-EDMs ${\tilde d_u}$ and  ${\tilde d_d}$ using the relations $g {\bar g}_0=- 0.37{\bar \theta}$ \cite{Witten} \footnote{Using updated results \cite{Yamanaka2017,Vries2015}:
    \begin{align}
       g \bar{g}_0 &= -0.2108 \bar{\theta} \, , \nonumber\\
       g \bar{g}_1 &= 46.24 \cdot 10^{-3} \bar{\theta} \, ,\nonumber
    \end{align}
we obtain practically the same  value of $S(\bar{\theta})$} 
and $g {\bar g}_0 = 0.8 \cdot10^{15}({\tilde d_u} +{\tilde d_d})$/cm,  $g {\bar g}_1 = 4 \cdot 10^{15}({\tilde d_u} - {\tilde d_d})$/cm  \cite{PR}:
  \begin{align}
  \label{Stheta} S( {\bar \theta})&= K_S  \,{\bar \theta} \, e\, \textrm{fm}^3,                    \\
  \label{SD}  S( {\tilde d})&= 10^4 K_S ( 0.50 \,{\tilde d}_u -  0.54 \,{\tilde  d}_d )\, e\, \textrm{fm}^2.        
  \end{align}   

\section{Nuclear electric dipole moments}

A nucleus with an intrinsic octupole deformation also has a small intrinsic EDM $D$ due to a difference between the proton and neutron distributions. 
A theoretical  estimate  in a two-liquid drop model is $D \approx 0.0004 A Z \beta_2 \beta_3 \approx 0.2$ e\,fm (see \cite{Spevak} and references therein).
Values of $D$ may be extracted from the half-lifes $T_{1/2}$ of the upper doublet state which are presented in \cite{nndt}. Using the formula for the electric dipole radiation we obtain  
\begin{equation}\label{dT}
\left(\frac{D}{\rm e\,fm}\right)^2 \sim
\frac{J+1}{J}\left(\frac{1.8 \cdot 10^{-12} {\rm s}}{T_{1/2}}\right) 
\left(\frac{100\,\mathrm{keV}}{|E_+-E_-|}\right)^3  .
\end{equation}
Using  half-lifes $T_{1/2}$ from \cite{nndt} we obtain intrinsic dipole moments $D(^{153}$Eu)=0.12 e\,fm,  $D(^{237}$Np)=0.013 e\,fm,  $D(^{227}$Ac)=0.061 e\,fm,
$D(^{161}$Dy)=0.072 e\,fm, $D(^{155}$Gd)=0.047 e\,fm. These numbers  may be compared to $D(^{225}$Ac)=0.25 e\,fm for an unstable $^{225}$Ac nucleus with the octupole deformation which was suggested in Refs. \cite{Auerbach,Spevak}. According to these papers the Schiff moment of the $^{225}$Ac  nucleus is 3 times larger than in $^{225}$Ra. 

 Similar to Eqs.  (\ref{n},\ref{Scol}), for the T,P-violating EDM  in the laboratory frame we obtain \cite{Auerbach,Spevak}  \footnote{A similar 
 $J^\pm$ 
doublet mixing mechanism produces huge enhancement of electron  EDM $d_e$  and  T,P-odd  interactions in polar molecules, such as ThO. Interaction of $d_e$ with molecular electric field produces the mixing coefficient $\alpha$ resulting in the orientation of  large intrinsic molecular EDM $D \sim e a_B$ along the molecular angular momentum ${\bf I}$, and we obtain  $d=2 \alpha \frac{I}{I+1} D \sim \alpha e a_B$  \cite{SushkovFlambaum}, where $a_B$ is the Bohr radius. As a result,  the T,P-violating molecular EDM $d$ exceeds electron EDM $d_e$ by 10 orders of magnitude. Experiment with ThO molecule \cite{TheEDM} gives a limit on electron EDM which is two orders of magnitude better then the limit from the atomic EDM measurement.}:
\begin{equation}\label{d}
 d=2 \alpha \frac{J}{J+1} D \,. 
\end{equation}
Here $\alpha$ is the mixing coefficient given by Eqs. (\ref{alpha},\ref{W}) and the intrinsic dipole moment  $D$ may be found from Eq. (\ref{dT}).
\section{Axion induced Schiff and electric dipole moments}

While the static nuclear EDM is completely shielded by electrons, the screening is incomplete if nuclear EDM is oscillating \cite{OscillatingEDM}.  Indeed, the  axion dark matter produces oscillating neutron EDM \cite{Graham}, oscillating nuclear EDM and oscillating nuclear Schiff moment \cite{Stadnik}. Oscillating  nuclear EDM and Schiff moment are enhanced  by the octupole mechanism. No different  nuclear calculations of the Schiff moment are needed. To obtain the results for the oscillating Schiff moment and EDM it is sufficient to replace the constant ${\bar \theta}$ by ${\bar \theta}(t)= a(t)/f_a$, where $f_a$ is the axion decay constant, $(a_0)^2= 2 \rho /(m_a)^2$, $\rho$ is the axion dark matter energy  density \cite{Graham,Stadnik}. 
Since an oscillating nuclear Schiff moment and oscillating nuclear EDM may be produced by the axion dark matter, corresponding measurements may be used to search for the dark matter. First results of such search have been published in Ref. \cite{nEDM}, where the oscillating neutron EDM and oscillating $^{199}$Hg Schiff moment have been measured.  Search for the effects produced by the an oscillating axion-induced Schiff moments in solid state materials is in progress \cite{Casper}. 

 Contribution of the nuclear EDM to  the EDM of heavy atoms is significantly smaller than the contribution of the Schiff moment since the oscillation frequency $\omega\approx m_a c^2/\hbar$  is small in comparison with the atomic transition frequency $\omega_{atomic}$.  As a result,  the electron screening still strongly suppresses nuclear EDM, $\sim \omega^2/(\omega_{atomic})^2$.
However, in molecules the partially screened nuclear EDM is $(M_{mol}/m_e)^2 \sim 10^6-10^8$ times larger than in atoms since nuclei in molecules move slowly and do not respond sufficiently fast to the variation of the nuclear EDM, i.e. nuclei do not produce efficient screening.  Moreover, in the case of the resonance between the frequency of the axion field oscillations and molecular transition frequency there may be an enormous  resonance enhancement of the oscillating nuclear EDM effect  \cite{OscillatingEDM} \footnote{The oscillating Schiff moment effect is also enhanced near the resonance.}. 
Therefore,  the collective nuclear EDM in nuclei with the octupole deformation given by Eq. (\ref{d}) may also be of interest. 

\section{Atomic and molecular electric dipole moments}

The electrostatic interaction between electrons and the nuclear Schiff moments produces electric dipole moments in atoms and molecules. 
The dependence of the atomic EDM on the nuclear charge $Z$ and nuclear radius is given by the following factor \cite{SFK} 
 \begin{equation}\label{KZA}
 K(Z,A) \approx Z^2 (\frac{a_B}{2ZR})^{2- 2 \gamma}\,,
 \end{equation}
 where $R=r_0 A^{1/3}$ is the nuclear radius, $r_0=1.2$ fm, $a_B$ is the Bohr radius, $\gamma=\sqrt{1 - (Z \alpha)^2}$.  It is convenient to normalize such factor to be equal to 1 for $^{225}$Ra since there are accurate relativistic many body numerical calculations for Ra atom and RaO molecule:
 \begin{equation}\label{KZ}
 K_Z \approx \frac{K(Z,A)}{K(88,225)}\,.
\end{equation}
Using calculations of Ra atom EDM from Refs. \cite{AtomicSchiff,AtomicSchiff2} one obtains the following estimate for an atomic EDM:
 \begin{eqnarray}\label{dS}
 d_a  \approx  -9 \cdot 10^{-17} K_Z  \frac{S}{|e|\, \textrm{fm}^3} |e| \, \textrm{cm}=\\
 K_S K_Z10^{-16} {\bar \theta}\, |e| \, \textrm{cm}\, .
  \end{eqnarray}
    The interaction constant $W_S$ for the effective T,P-violating interaction in molecules is defined by the following expression:
 \begin{equation}\label{WSdefinition}   
   W_{T,P}=W_S\frac{S}{J} {\bf J \cdot n}\,,  
 \end{equation}    
where ${\bf J}$ is the nuclear spin, ${\bf n}$ is  the  unit vector along the molecular axis in linear molecules. 
Using  calculation of $W_S$ for RaO from  Refs. \cite{RaO,RaOTitov}  we obtain an estimate
\begin{equation}\label{WSZ}   
 W_S= 45192 K_Z
 \end{equation}    
 in atomic units (here a.u.=$e/a_B^4$). Substitution of the Schiff moment (\ref{Stheta}) to the energy shift $W_{T,P}=W_S\frac{S}{J} {\bf J \cdot n}$ gives for the fully polarised molecule the energy difference between the $J_z=J$ and $J_z=-J$ states:
 \begin{equation}\label{Wtheta} 
 2 W_S S=0.5 \cdot 10^7  K_S K_Z {\bar \theta} \, h\, \textrm{Hz},
 \end{equation}    
  where $h$ is the Plank constant.  Thus, we consider molecules where the shift $\sim 10^7 {\bar \theta} \, \textrm{Hz}$. With the current limit $| {\bar \theta} | < 10^{-10}$ the maximal shift is $10^{-3}$ Hz. The measured shift in the 1991 TlF experiment \cite{TlFexperiment} was $(-1.3 \pm 2.2) \cdot 10^{-4}$ Hz, i.e. such accuracy is already sufficient.  
  It is expected that new generation of molecular experiments will improve this accuracy by several orders of magnitude \cite{NewTlF}.  Therefore, we may expect a very significant  
 improvement of  the current limit  $| {\bar \theta} | < 10^{-10}$ and also improvement of the limits on other fundamental parameters of the CP-violation theories such as the strength of T,P-violating potential $\eta$, the $\pi NN$ interaction constants ${\bar g}$ and the quark chromo-EDMs ${\tilde d}$.

\section{Nuclear Schiff moment, atomic EDM and molecular spin-axis constant for $^{237}\textrm{Np}$,  $^{153}\textrm{Eu}$, $^{235}\textrm{U}$ and $^{227}\textrm{Ac}$}
  
   Unfortunately, the nuclei with intrinsic octupole deformation and non-zero spin suggested in Refs.\cite{Auerbach,Spevak} have a short lifetime. Several experimental groups have considered experiments with  $^{225}$Ra and $^{223}$Rn       \cite{RaEDM,RaEDM2,RnEDM}. 
    The only published EDM measurements  \cite{RaEDM,RaEDM2}  have been done for $^{225}$Ra which has 15 days half-life. In spite of the Schiff moment enhancement the $^{225}$Ra  EDM measurement has not reached yet the sensitivity to the T,P-odd interaction  (Eq.~ (\ref{eta})) comparable to the Hg EDM experiment \cite{HgEDM}. The experiments continue, however,  the instability of $^{225}$Ra and a relatively small number of atoms available  may be a problem. In Ref. \cite{Th} the    
nuclear Schiff moment of $^{229}$Th nucleus has been estimated since this nucleus has a much longer lifetime (7917 years).  Spectrum of even-even nucleus $^{228}$Th presented in the database \cite{nndt} indicates the static octupole deformation. However, adding a neutron to this nucleus and forming    $^{229}$Th seems to blur  features of the rotational spectrum for the  octupole deformation
\footnote{One of the problems here is that we need a nucleus with a non-zero spin, while   octupole deformations 
are mostly studied
in  even-even nuclei with zero spin - see e.g \cite{Robledo2013,Nomura2015,Afanasjev2016,Sm152}. 
However, based on nuclear rotational spectra discussed in the Section~\ref{sec:exp-octupole} we observe that adding a proton to an 
octupole deformed even-even nucleus increases the chance that the octupole deformation survives in the odd-even nucleus with half-integer spin. Whereas adding a neutron seems to blur typical features of octupole deformation. 
This looks natural since an additional proton increases the Coulomb repulsion and tends to stretch the nucleus.}. The enhanced Schiff moment in this case may be produced by a soft octupole vibration mode. 

So far no stable  nuclei, which have  noticeable natural abundance and octupole deformation, have been suggested for the EDM experiments. In Section~\ref{sec:exp-octupole} we
aim to identify such nuclei or  nuclei with a soft octupole vibration mode (which is a precursor of the octupole deformation)
by their measured rotational excitation spectra.
We should find nuclei with a low lying excited state of opposite parity and the same angular momentum as the ground state. 
Rotational bands built on both of these states should have close values of the moment of inertia indicating collectivity.
The candidates include  stable nuclei $^{153}$Eu, $^{161}$Dy, $^{163}$Dy, $^{155}$Gd, and long lifetime nuclei   $^{235}$U, $^{237}$Np, $^{233}$U, $^{229}$Th, $^{153}$Sm, $^{165}$Er, $^{225}$Ac, $^{227}$Ac, $^{231}$Pa, $^{239}$Pu. In this paper we concentrate on the most attractive  cases of  $^{153}$Eu, $^{235}$U, $^{237}$Np and $^{227}$Ac. We start from a simple estimate of the Schiff moments in these nuclei assuming that they have intrinsic octupole deformation. A detailed  discussion of experimental evidence for the octupole deformation in these nuclei is presented in the section  \ref{sec:exp-octupole}.

\subsection{$^{237}$Np}

The half-life of $^{237}$Np is 2.14 million years  and  macroscopic quantities are produced in nuclear reactors. 
Experimental nuclear excitation spectra in this nucleus satisfy criteria for the octupole deformation, the interval between opposite parity levels which are mixed by the T,P-odd interaction is $E(\frac{5}{2}^-) - E(\frac{5}{2}^+)=59.5$ keV.  This nucleus has proton above  $^{236}$U nucleus and deformation parameters (interpolated between $^{234}$U and $^{238}$U from Ref.\cite{Afanasjev2016} ) $\expect{\beta_3^2 } =(0.12)^2$  and $\beta_2$=0.26. Eqs.(\ref{San} - \ref{SD} ) give us strongly enhanced values  of the $^{237}$Np Schiff moment ($K_S$=6.5, i.e. the Schiff moment is 6.5 times larger than that in $^{225}$Ra and  3 orders of magnitude larger than that in spherical nuclei):
 \begin{align}\label{Npschiff}
 S(^{237}{\rm Np},\eta)&  \approx 2 \cdot 10^{-5}  e \,\eta \, \textrm{fm}^3 \, ,\\
 S(^{237}{\rm Np},g) & \approx  6 (  - 2.6 g {\bar g}_0 +  12.9 g {\bar g}_1 -6.9 g {\bar g}_2)\, e\, \textrm{fm}^3\, ,\\
 S(^{237}{\rm Np}, {\bar \theta})& \approx 6 \,{\bar \theta} \, e\, \textrm{fm}^3 \ ,\\
 S(^{237}{\rm Np},{\tilde d})& \approx 6 ( 0.50 \,{\tilde d}_u -  0.54 \,{\tilde  d}_d )\, e\, \textrm{fm}^2 \, .
 \end{align}
 The electron scaling factor for Np is $K_Z$=1.4 and the atomic EDM is
 \begin{align}\label{Npd}
 d_a ( ^{237}{\rm Np})& \approx  -  1.2  \cdot 10^{-16}   \frac{S}{|e|\, \textrm{fm}^3} |e| \, \textrm{cm} \\
 &\approx  -  0.8  \cdot 10^{-15} {\bar \theta}\, |e| \, \textrm{cm}\ .
 \end{align}   
 The interaction constant $W_S$ for the effective T,P-violating interaction in 
 molecules containing $^{237}$Np is:
 \begin{equation}\label{Npws}   
 W_S \, \approx \,63000 \,\,\textrm{a.u.} \ .
 \end{equation}   
 The energy shift is
 \begin{equation}\label{Nptheta} 
 2 W_S S=5 \cdot 10^7  {\bar \theta} \, h\, \textrm{Hz}.
 \end{equation}   

\subsection{$^{153}$Eu}

$^{153}$Eu  is stable with 52\% natural abundance, experimental nuclear excitation spectra in this nucleus satisfy all criteria for the octupole deformation, the interval between opposite parity levels which are mixed by the T,P-odd interaction is small, $E(\frac{5}{2}^-) - E(\frac{5}{2}^+)=97.4$ keV.  This nucleus has a proton above  $^{152}$Sm  which according to Ref.  \cite{Sm152} has octupole deformation. If we for an estimate  assume the deformation parameters from Ref.   \cite{Sm152},   $\expect{\beta_3^2 } =(0.15)^2$ (calculated for $^{152}$Sm ) and $\beta_2$=0.31 (experimental value  for $^{152}$Sm ), Eq.(\ref{San}) gives us strongly enhanced values  of the $^{153}$Eu Schiff moment ($K_S$=3.7):
\begin{align}\label{Eu}
 S(^{153}{\rm Eu},\eta)&  \approx 1.1 \cdot 10^{-5}  e \,\eta \, \textrm{fm}^3,\\
 S(^{153}{\rm Eu},g)& \approx  3.7 (  - 2.6 g {\bar g}_0 +  12.9 g {\bar g}_1 -6.9 g {\bar g}_2)\, e\, \textrm{fm}^3,\\
 S(^{153}{\rm Eu}, {\bar \theta})&\approx  3 .7 \,{\bar \theta} \, e\, \textrm{fm}^3,  \\
 S( ^{153}{\rm Eu},{\tilde d})&\approx  3.7 ( 0.50 \,{\tilde d}_u -  0.54 \,{\tilde  d}_d )\, e\, \textrm{fm}^2\, .  
 \end{align}
 The electron scaling factor for Eu is $K_Z$=0.22, therefore,  the atomic EDM and molecular EDM are comparable to that for $^{225}$Ra ($K_S K_Z=0.8$): 
 \begin{eqnarray}\label{EudS}
 d_a ( ^{153}{\rm Eu}) \approx  -2 \cdot 10^{-17}   \frac{S}{|e|\, \textrm{fm}^3} |e| \, \textrm{cm}\,\,\,\, \\
\approx  -  0.8  \cdot 10^{-16} {\bar \theta}\, |e| \, \textrm{cm}\,\,.\,\,\,
 \end{eqnarray}   
 The EDM is slightly bigger in the $^{153}$Eu$^{3+}$ ion which has zero electron angular momentum and may be convenient for Schiff moment and nuclear EDM measurements. 
 
 The interaction constant $W_S$ for the effective T,P-violating interaction in $^{153}$Eu-containing molecules such as $^{153}$Eu\,N is :
 \begin{equation}\label{EuWSZ}   
 W_S \, \approx \,10500 \,\,\textrm{a.u.}
 \end{equation}    
 The energy shift is
 \begin{equation}\label{WTP} 
 2 W_S S=0.4 \cdot 10^7  {\bar \theta} \, h\, \textrm{Hz}.
 \end{equation}    
 Here it is also worth looking for molecules with zero electron angular momentum in the ground or metastable excited state.  
The Schiff moment can also be measured in ferroelectric solids containing $^{153}$Eu. An advantage here is the strong internal electric field acting on  the $^{153}$Eu atom and a macroscopic number of atoms coherently contributing to the observable effect. Currently  the experiment is in progress with a $^{207}$Pb-containing ferroelectric solid where the Schiff moment is not enhanced \cite{Casper}. Eu-containing ferroelectric solids are used to measure  electron EDM \cite{Lamoreaux}. Note that one may also look for a possibility to measure  Schiff moments of actinide nuclei  $^{235}$U, $^{237}$Np and $^{237}$Ac in ferroelectric solids.

\subsection{$^{235}$U}

$^{235}$U  is practically stable (half life 0.7 billion years), with 0.75\% natural abundance. The interval between opposite parity levels which are mixed by the T,P-odd interaction is $E(\frac{7}{2}^+) - E(\frac{7}{2}^-)=81.7$ keV.  This nucleus has a neutron above  $^{234}$U nucleus which according to Ref.~\cite{Afanasjev2016} has octupole deformation with  $\expect{\beta_3^2 } =(0.17)^2$  and $\beta_2$=0.25.  However, experimental nuclear excitation spectra of $^{235}$U  do not show  parity doublets,
see Fig.~\ref{fig:234U}.
Instead, there are opposite parity rotational bands which start from lower values of the minimal angular momenta. The octupole deformation in $^{234}$U means that $^{235}$U should  have  a soft octupole vibration mode. For an estimate we can use   
$\expect{\beta_3^2 } =(0.1)^2$.  Using also $\beta_2$=0.25 we obtain $K_S=3$ and $^{235}$U Schiff moment:
\begin{align}\label{Uschiff}
 S(^{235}{\rm U},\eta) & \approx 1 \cdot 10^{-5}  e \,\eta \, \textrm{fm}^3\, ,\\
 S(^{235}{\rm U},g) &\approx  3 (  - 2.6 g {\bar g}_0 +  12.9 g {\bar g}_1 -6.9 g {\bar g}_2)\, e\, \textrm{fm}^3\, ,\\
 S(^{235}{\rm U}, {\bar \theta})&\approx  3  \,{\bar \theta} \, e\, \textrm{fm}^3 \, ,\\
 S( ^{235}{\rm U},{\tilde d})&\approx  3 ( 0.50 \,{\tilde d}_u -  0.54 \,{\tilde  d}_d )\, e\, \textrm{fm}^2\, .  
 \end{align}
 We should note that these estimates are less reliable than that for other nuclei in this paper
  since the magnitude of the Schiff moment due the soft octupole mode  has not been  calculated yet using a microscopic approach as it was done for nuclei with the static octupole.
  
 The electron scaling factor for the U atom  is $K_Z$=1.3 and the atomic EDM is
 \begin{align}\label{Ud}
 d_a ( ^{235}{\rm U})& \approx  - 1.2 \cdot 10^{-16}   \frac{S}{|e|\, \textrm{fm}^3} |e| \, \textrm{cm} \\
&\approx  -  0.4 \cdot 10^{-15} {\bar \theta}\, |e| \, \textrm{cm}\ .
 \end{align}   
 The interaction constant $W_S$ for the effective T,P-violating interaction in molecules containing $^{235}$U is:
 \begin{equation}\label{Uws}   
 W_S \, \approx \,59000 \,\,\textrm{a.u.}
 \end{equation}    
 The energy shift is
 \begin{equation}\label{Utheta} 
 2 W_S S=2 \cdot 10^7  {\bar \theta} \, h\, \textrm{Hz},
 \end{equation}  

\subsection{$\Ac$}
$\Ac$ with a half-life of 21.8 years is shorter lived than the other examples but is readily available as it is produced commercially for cancer treatment. 
$\Ac$  is a product of $^{225}$U decay chain. It is also produced in nuclear reactors 
by neutron capture of $^{226}$Ra.
Experimental nuclear excitation spectra in this nucleus satisfy criteria for the octupole deformation, the interval between opposite parity levels which are mixed by the T,P-odd interaction is $E(\frac{3}{2}^+) - E(\frac{3}{2}^-)=27.37$ keV.  
This nucleus has a proton above  $^{226}$Ra which according to Ref.~\cite{Afanasjev2016} has an octupole deformation with  $\expect{\beta_3^2 } =(0.134)^2$  and $\beta_2$=0.197.
 This gives $K_S=10$  and strongly enhanced values  of the $\Ac$ Schiff moment:
\begin{align}\label{Acschiff}
 S(\Ac,\eta)&  \approx 3 \cdot 10^{-5}  e \,\eta \, \textrm{fm}^3 \, ,\\
 S(\Ac,g) & \approx  10 (  - 2.6 g {\bar g}_0 +  12.9 g {\bar g}_1 -6.9 g {\bar g}_2)\, e\, \textrm{fm}^3\, ,\\
 S(\Ac, {\bar \theta})& \approx 10 \,{\bar \theta} \, e\, \textrm{fm}^3 \ ,\\
 S(\Ac,{\tilde d})& \approx 10 ( 0.50 \,{\tilde d}_u -  0.54 \,{\tilde  d}_d )\, e\, \textrm{fm}^2 \, .
 \end{align}
 The electron scaling factor for Ac is $K_Z$=1.07 and the atomic EDM is
 \begin{align}\label{Acd}
 d_a (\Ac)& \approx  -  1.0  \cdot 10^{-16}   \frac{S}{|e|\, \textrm{fm}^3} |e| \, \textrm{cm} \\
 &\approx  -  1.0  \cdot 10^{-15} {\bar \theta}\, |e| \, \textrm{cm}\ .
 \end{align}   
 An interesting case here may be Ac$^+$ ion which has an electronic structure similar to the Ra atom, zero electron angular momentum and consequently a long nuclear spin coherence time. 
A method to measure the EDM of ions has been suggested and implemented in Ref.~\cite{Cornell}. 
 
 The molecule Ac\,F has an electronic structure similar to Ra\,O which has zero electron angular momentum in the  ground state, $^1 \Sigma$ term. The molecule Ac\,O\,H may be even better since similar to  Ra\,O\,H$^+$ considered in Ref. \cite{MOH+}  this kind of molecules have doublets  of close opposite parity levels and may be easily polarised by an external electric field.
    
    The interaction constant $W_S$ for the effective T,P-violating interaction in $^{227}$Ac-containing  molecules is :
 \begin{equation}\label{Acws}   
 W_S \, \approx \,46000 \,\,\textrm{a.u.} \ .
 \end{equation}   
 The energy shift is
 \begin{equation}\label{Actheta} 
 2 W_S S=5 \cdot 10^7  {\bar \theta} \, h\, \textrm{Hz}.
 \end{equation}   
Using opportunity we would like to note that Pa$^3+$ ion and PaN molecule have electronic structure similar to Ra atom and ThO molecule and  may have large effects of $T,P$-violation
(see e.g. Refs. \cite{Auerbach,Spevak} where the calculation of Schiff moment in the unstable nucleus $^{229}$Pa is performed).  

The reader should be reminded
that all numbers presented above are results of the extrapolation from the accurate many-body calculations performed for the  $^{225}$Ra nucleus, Ra atom and Ra\,O molecule and should be treated as order of magnitude estimates. 


\section{Experimental evidence for intrinsic octupole deformation}
\label{sec:exp-octupole}

In atomic physics it is quite obvious that a molecule can have a deformed electron distribution with various kinds of multipole moments.
The most simple example, which may serve as an analogue to the nuclear case, is a diatomic molecule with different nuclei.
The different positive charges of the nuclei produce a Coulomb field that imprints an octupole deformation on the electron density. 
This distribution is denoted as the body fixed density or intrinsic density.
On the other hand if the ground state of the dimer has total spin $I^\pi=0^+$ it has  a spherical charge distribution. 
The intrinsic pear-shaped density reveals itself when exciting the molecule. 
A rotational spectrum with $I^\pi = 0^+, 1^-, 2^+, 3^-, 4^+, \cdots$, where the excitation energies of positive and negative states  follow $E^{I^\pi}\!\!-E^{I^\pi\!=0^+}=I(I+1)/2\theta_{mol}$ with the same moment of inertia 
$\theta_{mol}$ for both parities, is a clear signature of a rotating octupole deformed system.
  
If one takes the two nuclei to be equal the negative parity states are missing in the rotational spectrum because the Coulomb field is symmetric under parity operation.

The situation is analogue in nuclei, except that there is no external field that enforces the octupole deformation and could be used to define a body fixed frame. 
Nuclei are self-bound and their intrinsic deformation is a result of the interactions among the nucleons.
Thus, it is not as obvious as in molecules to predict when intrinsic deformations occur.
But one can inspect the energy spectrum of the nuclear eigenstates and transitions to see if the picture of a rotating intrinsically deformed system is applicable.

Mean field models, which provide a simple approximation for the nuclear many-body state, are helpful in setting up the concept of intrinsic deformation. 
In the Hartree-Fock (HF) approximation one minimizes the expectation value of the nuclear Hamiltonian with respect to variations of the single-particle states comprising a trial Slater determinant. 
In the Hartree-Fock-Bogoliubov (HFB) approximation the trial many-body state includes pairing correlations by means of a Bogoliubov ansatz. 
In the following we refer to both as mean field states.

Minimization of the energy may lead to a mean field state that breaks the symmetries of the Hamiltonian. 
In many cases one obtains for example a single Slater determinant $\ket{Q}$ with a one-body density which is not spherically symmetric although the ground state of the nuclear system has zero total angular momentum, $J^\pi=0^+$.
One calls such HF or HFB states intrinsic states and the nucleus they describe as intrinsically deformed.

The rotational symmetry and parity of the nuclear system can be restored by projecting out of an intrinsic mean field state $\ket{Q}$,  a many-body state which is eigenstate of $\op{J}$, $\op{J}_z$, and the parity operation $\op{\Pi}$, 
\begin{align}\label{eq:proj}
	\ket{J^\pm M,n=0} &\approx \op{P}^{\,J}_{MK=0}\ket{\pm;Q} \\
	\mathrm{with}\ \ &\ket{\pm;Q}=\frac{1}{2} \Big( \ket{Q}\pm\op{\Pi}\ket{Q} \Big)\ .
\end{align}
If the concept of a rotating deformed system is valid, these projected states can be regarded as good approximations of the exact eigenstates of the nuclear Hamiltonian in Eq.~\eqref{eq:eigen}.
The energies follow in that case the pattern of a rotational band with the moment of inertia $\theta^\pi$.
\begin{align}\label{eq:E}
E^{J^\pi}=E^{J^\pi}_0 +\frac{J(J+1)}{2\theta^\pi}
\end{align}
For a well developed intrinsic octupole deformation or a soft octupole vibration the moments of inertia $\theta^+$ and $\theta^-$ should be the same for positive and negative parity.
If measured energies and transition rates are well reproduced by the concept of a rotating intrinsic state the picture of an  intrinsic density makes sense. 

One expects that a parity pair which can be approximated by a spin and parity projection of the same octupole deformed intrinsic state $\ket{Q}$
yields large values for the matrix element of $S$ due to collectivity.
We searched for stable or long lived isotopes throughout the nuclear chart which match the above criteria for large octupole moments in the nuclear charge distribution. We identified $^{237}$Np, $^{153}$Eu, $^{235}$U and $^{227}$Ac as possible candidates for atomic experiments searching for P,T-violation.


Let us first consider the even-even nuclei which have one nucleon less than the proposed isotopes.
For all four cases the even-even nuclei have been investigated in mean-field models \cite{Afanasjev2016,Sm152} and 
intrinsic octupole moments have been found as the spectra suggest. 
Their strengths however depend on the energy functional chosen. 
Beyond mean-field models which project on angular momentum, parity and particle number, and include mixing of configurations with different intrinsic quadrupole and octupole moments by means of the Generator Coordinate Method (GCM) are available for $^{224}$Ra \cite{Robledo2013} and for $^{152}$Sm \cite{Nomura2015}.

In this section, however, we base our investigations only on experimental spectra.
In the following all experimental data are extracted from Ref.~\cite{nndt}.

\subsection{$^{236}$U and $^{237}$Np}
Let us consider a nucleus with an even number of protons and neutrons (spin saturated) and an intrinsic axial deformation around the $z$-axis $(K=0)$.
\begin{figure}[ht]
\includegraphics[width=0.45\textwidth]{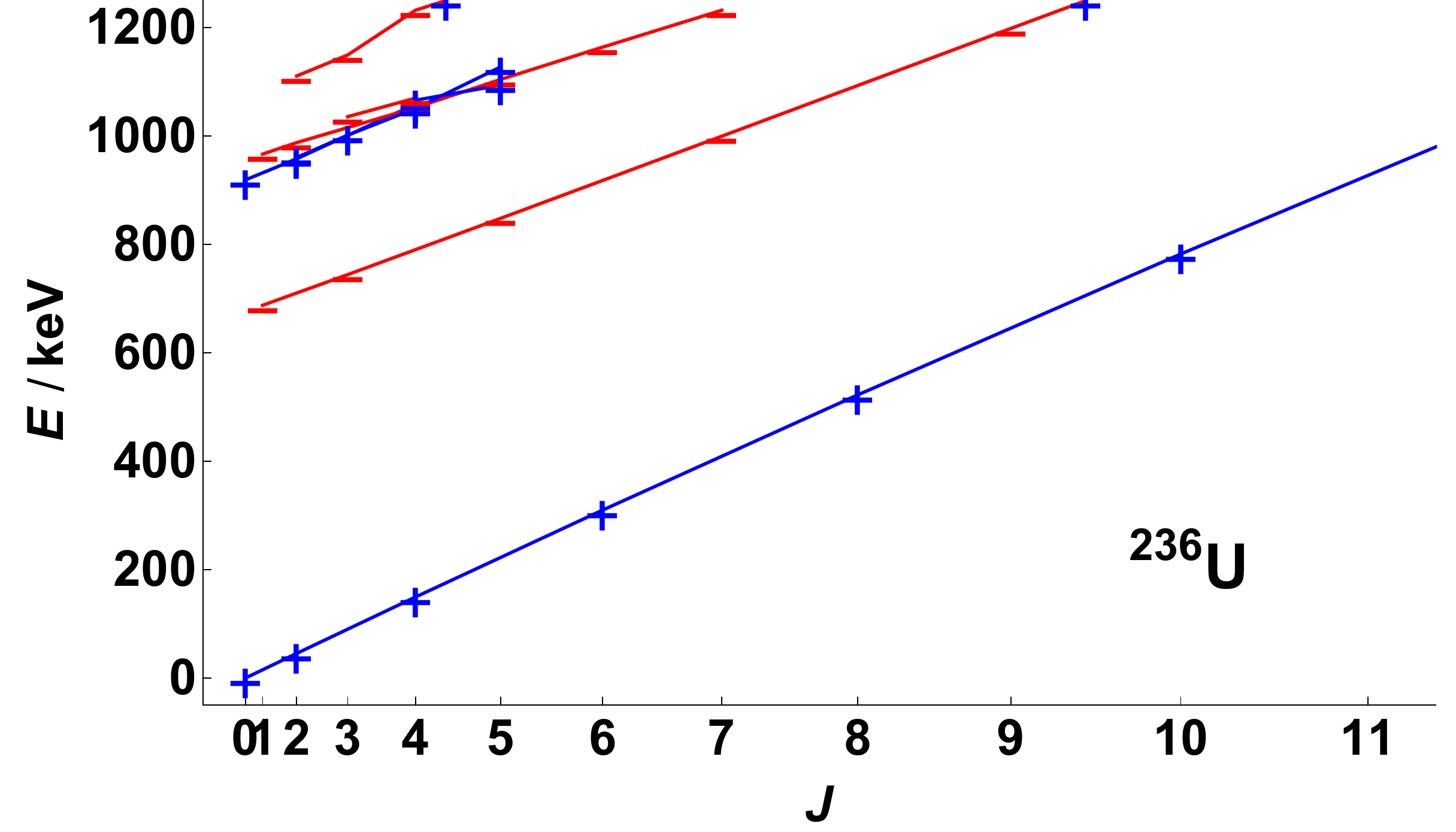}
\caption{\label{fig:236U}
Energy levels of $^{236}$U as function of $J(J+1)$ for all experimentally observed states below 1200~keV. Positive and negative parities are indicated by {\color{blue} \textbf{+}} and {\color{red} \textbf{ --} }, respectively. Straight lines indicate rotational bands with $E=E_0+J(J+1)/2{\Theta}$.\\
}
\end{figure}
\begin{figure}[h]
\includegraphics[width=0.45\textwidth]{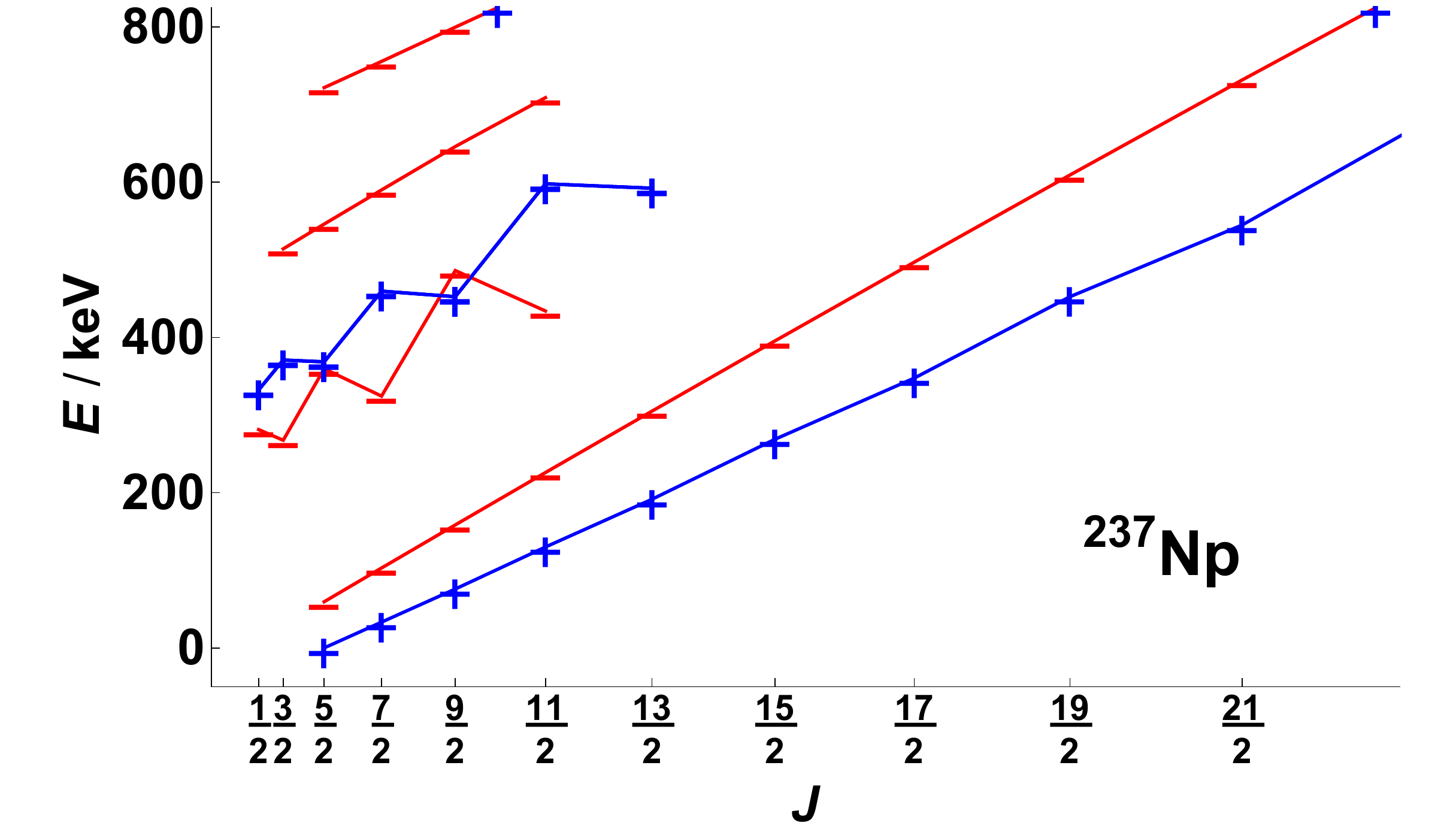}
\caption{\label{fig:237Np}
Energy levels of $^{237}$Np as function of $J(J+1)$ for all experimentally observed states below 800~keV. Positive and negative parities are indicated by {\color{blue} \bf +} and {\color{red} \bf -- }, respectively. Straight lines indicate rotational bands with $E=E_0+J(J+1)/2{\Theta}$. 
}
\end{figure}
If the intrinsic state $\ket{Q}$ is not invariant under parity projection one expects after projection on positive parity a rotational band $J^\pi = 0^+, 2^+, 4^+, \cdots$ where the energies grow proportional to $J(J+1)$.
This is seen
in Fig.~\ref{fig:236U} where the energy levels and spin-parity assignments $J^\pi$ of $^{236}$U, a nucleus with even proton ($Z=92$) and even neutron number ($N=144$),  are displayed as function of $J(J+1)$.
Projection on negative parity will lead to a rotational band with $J^\pi = 1^-, 3^-, 5^-, \cdots$.
These two rotational bands are clearly visible in Fig.~\ref{fig:236U}. 

But the negative parity band is about 650~keV higher than the positive parity band.
This indicates that the octupole deformation is of softer nature. For negative parity the collective wave function for the collective degree of freedom $\beta_3$ gets a node for $\beta_3=0$, while for positive parity it is symmetric in going from $\beta_3$  to $-\beta_3$. This additional node causes collective kinetic energy which accounts for the difference in energy between the positive and negative parity rotational bands.
The moments of inertia are similar, $\theta^-$ being somewhat larger than $\theta^+$, which is in accord with the idea that the additional node stretches the whole configuration somewhat compared to the symmetric case.
For more details on this effect see for example Refs.~\cite{Bernard2016,Fu2018,Bucher2017,Lica2018}.

By adding a proton to $^{236}$U one obtains $^{237}$Np (Z=93, N=144) the spectrum of which is displayed in Fig.~\ref{fig:237Np}.
One observes two rotational bands with equal total spins and opposite parities which is a typical footprint of intrinsic octupole deformation.
The energy difference between the lowest 
$J^\pi=5/2^+$ and $J^\pi=5/2^-$ is only 60~KeV.

\subsection{$^{152}$Sm and $^{153}$Eu}
\begin{figure}[ht]
\includegraphics[width=0.4\textwidth]{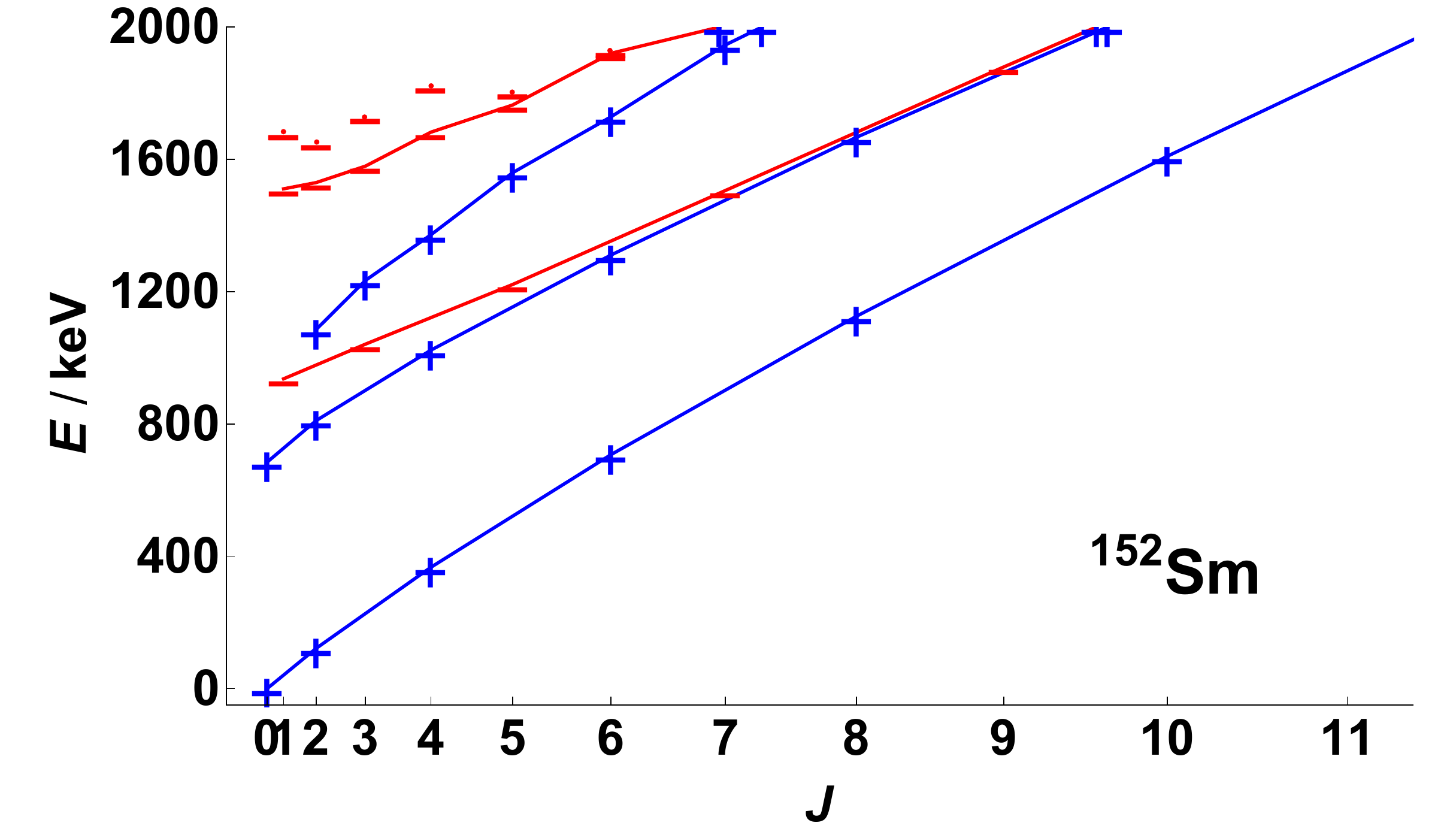}
\caption{\label{fig:152Sm}
Energy levels of $^{152}$Sm as function of $J(J+1)$ for experimentally observed states below 2000~keV. Positive and negative parities are indicated by {\color{blue} \bf +} and {\color{red} \bf -- }, respectively. Straight lines indicate rotational bands with $E=E_0+J(J+1)/2{\Theta}$.
}
\end{figure}

Another example which shows typical signs for an intrinsic octupole deformation is $^{152}$Sm. 
The spectrum displayed in Fig.~\ref{fig:152Sm} shows a rotational ground state band $J^\pi = 0^+_1, 2^+_1, 4^+_1, \cdots$ and its octupole partner, the negative parity band $J^\pi = 1^-_1, 3^-_1, 5^-_1, \cdots$, about 930~keV higher, with a somewhat larger but still similar moment of inertia.
 
In so far the situation is similar to $^{236}$U, but $^{132}$Sm has in addition two excited rotational bands with positive parity,
$J^\pi = 0^+_2, 2^+_2, 4^+_2, \cdots$ and $J^\pi = 2^+_3, 4^+_2, 5^+_1, \cdots$.
These two indicate a triaxially deformed intrinsic state with $K=0$ and $K=2$, respectively.
Therefore one expects these two bands to originate from an intrinsic state which is triaxial and reflection symmetric.
There is no negative parity band with similar moment of inertia that could be the octupole partner to the triaxially deformed band. 
\begin{figure}[h]
\includegraphics[width=0.4\textwidth]{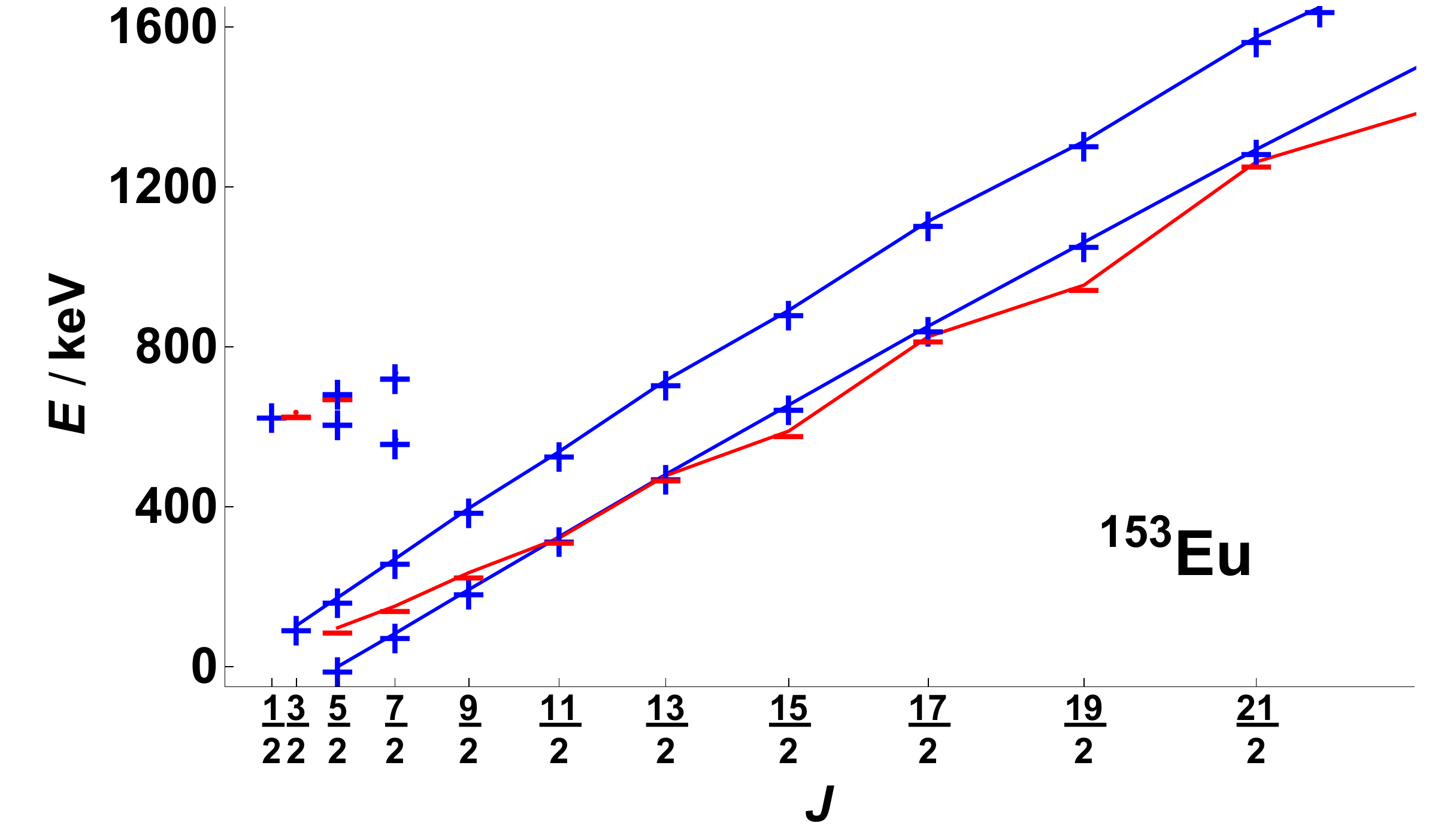}
\caption{\label{fig:153Eu}
Energy levels of $^{153}$Eu as function $J(J+1)$ for experimentally observed states below 1600~keV. States with ($E>700$~keV, $J<9/2$) are not included as they don't show rotational behavior. Positive and negative parities are indicated by {\color{blue} \bf +} and {\color{red} \bf -- }, respectively. Straight lines indicate rotational bands with $E=E_0+J(J+1)/2{\Theta}$. }
\end{figure}

Adding a proton leads to $^{153}$Eu (Z=63, N=90) with a spectrum shown in Fig.~\ref{fig:153Eu}. 
Like in $^{237}$Np one sees the parity partners close to each other indicating an intrinsic octupole deformation. 
The positive and negative parity bands $J^\pi=(5/2^+,5/2^-), (7/2^+,7/2^-), \cdots$ even merge at larger values of $J$. 

There is no obvious negative parity band as partner of the excited $J^\pi=3/2^+, 5/2^+, 7/2^+, \cdots$ band starting at 103~keV.

\subsection{$^{234}$U and $^{235}$U}
\begin{figure}[ht]
\includegraphics[width=0.4\textwidth]{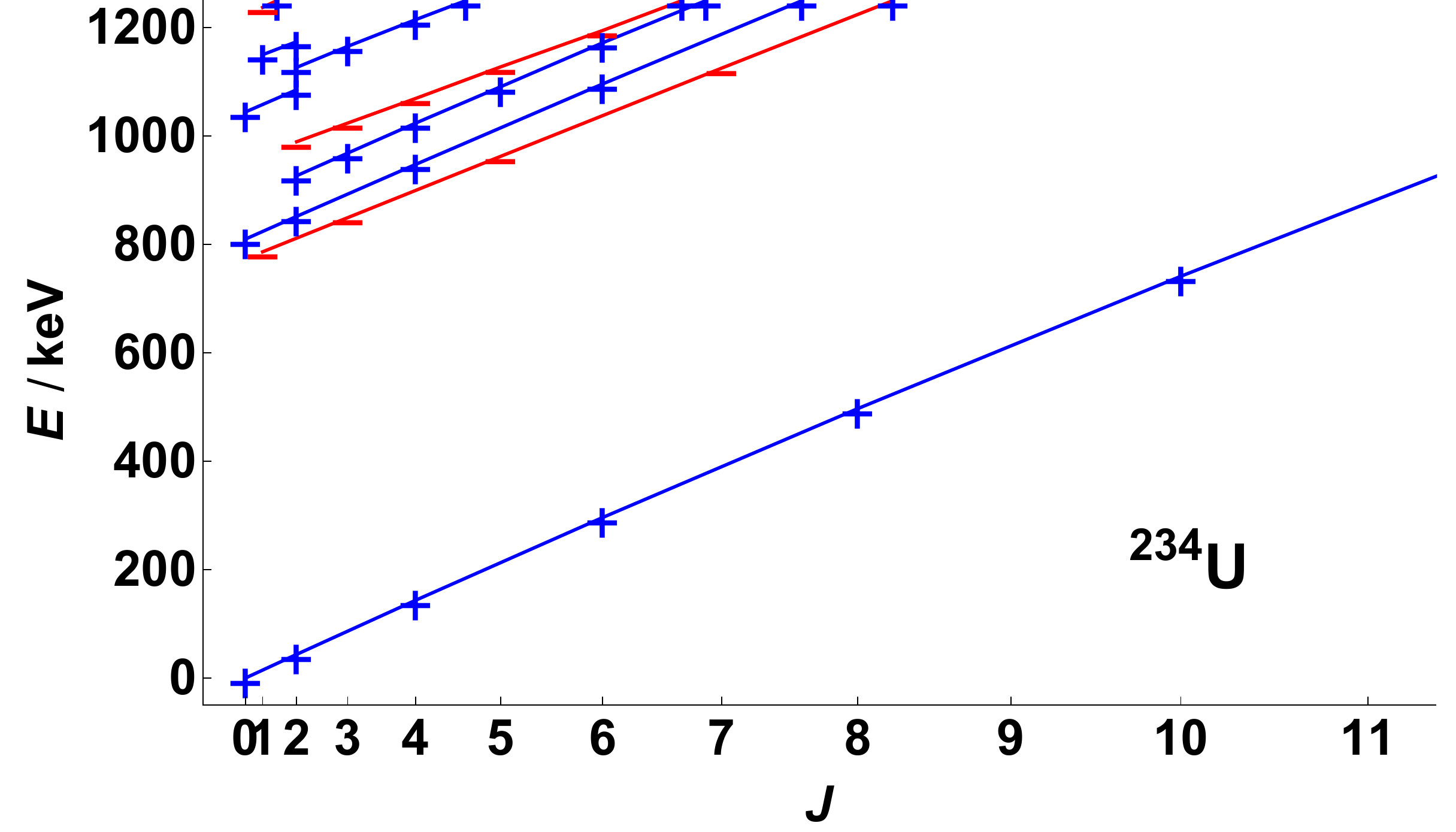}
\caption{\label{fig:234U}
Energy levels of $^{234}$U as function of $J(J+1)$ for experimentally observed states below 1200~KeV. Positive and negative parities are indicated by {\color{blue} \bf +} and {\color{red} \bf -- }, respectively. Straight lines indicate rotational bands with $E=E_0+J(J+1)/2{\Theta}$.
}
\end{figure}

The third example is $^{235}$U where a neutron is added to 
$^{234}$U.
 The spectrum of $^{234}$U (Fig.~\ref{fig:234U}) is similar to the one of $^{236}$U but in $^{234}$U further rotational bands come close to the $J^\pi = 1^-, 3^-, 5^-, \cdots$ band.
\begin{figure}[h]
\includegraphics[width=0.4\textwidth]{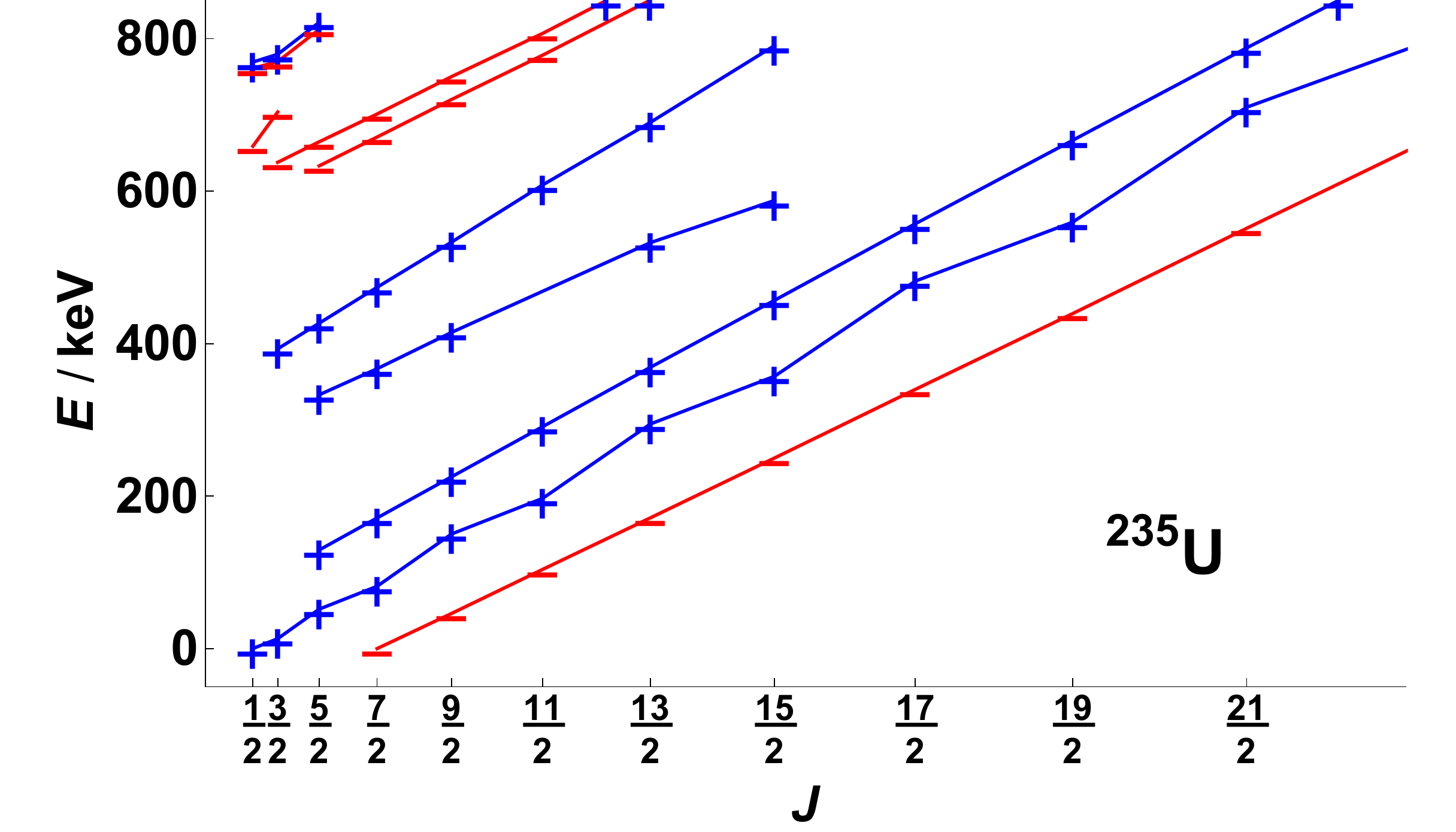}
\caption{\label{fig:235U}
Energy levels of $^{235}$U as function $J(J+1)$ for experimentally observed states below 800~keV. Positive and negative parities are indicated by {\color{blue} \bf +} and {\color{red} \bf -- }, respectively. Straight lines indicate rotational bands with $E=E_0+J(J+1)/2{\Theta}$. }
\end{figure}

As can be seen in Fig.~\ref{fig:235U} adding a neutron leads in $^{235}$U to a more complex spectrum than in the other cases.
Now the ground state band has $J^\pi=7/2^-, 9/2^-, 11/2^-,\cdots$ and there are two positive parity bands just above with lowest spin $1/2^+$ and $5/2^+$, respectively.
The first excited $J^\pi=1/2^+_1, 3/2^+_1, 5/2^+_1,\cdots$ positive parity band shows some staggering due to Coriolis effects. 
As the lowest spin is $1/2^+$ it can not be the octupole partner of the groundstate band.
The second positive parity band  $J^\pi=5/2^+_2, 7/2^+_2, 9/2^+_2,\cdots$ can also not be the ideal octupole partner of the groundstate band as it starts with $J^\pi=5/2^+_2$. Thus the extra neutron has blurred the picture and might have destroyed to a certain extend octupole properties.

Nevertheless the energy difference between the ground state $J^\pi=7/2^-_1$ and the excited $J^\pi=7/2^+_1$ state is only 
81.7 keV and parity violating matrix elements are not necessarily smaller than in other cases. 

\subsection{$^{226}$Ra and $^{227}$Ac}

\begin{figure}[h]
\includegraphics[width=0.4\textwidth]{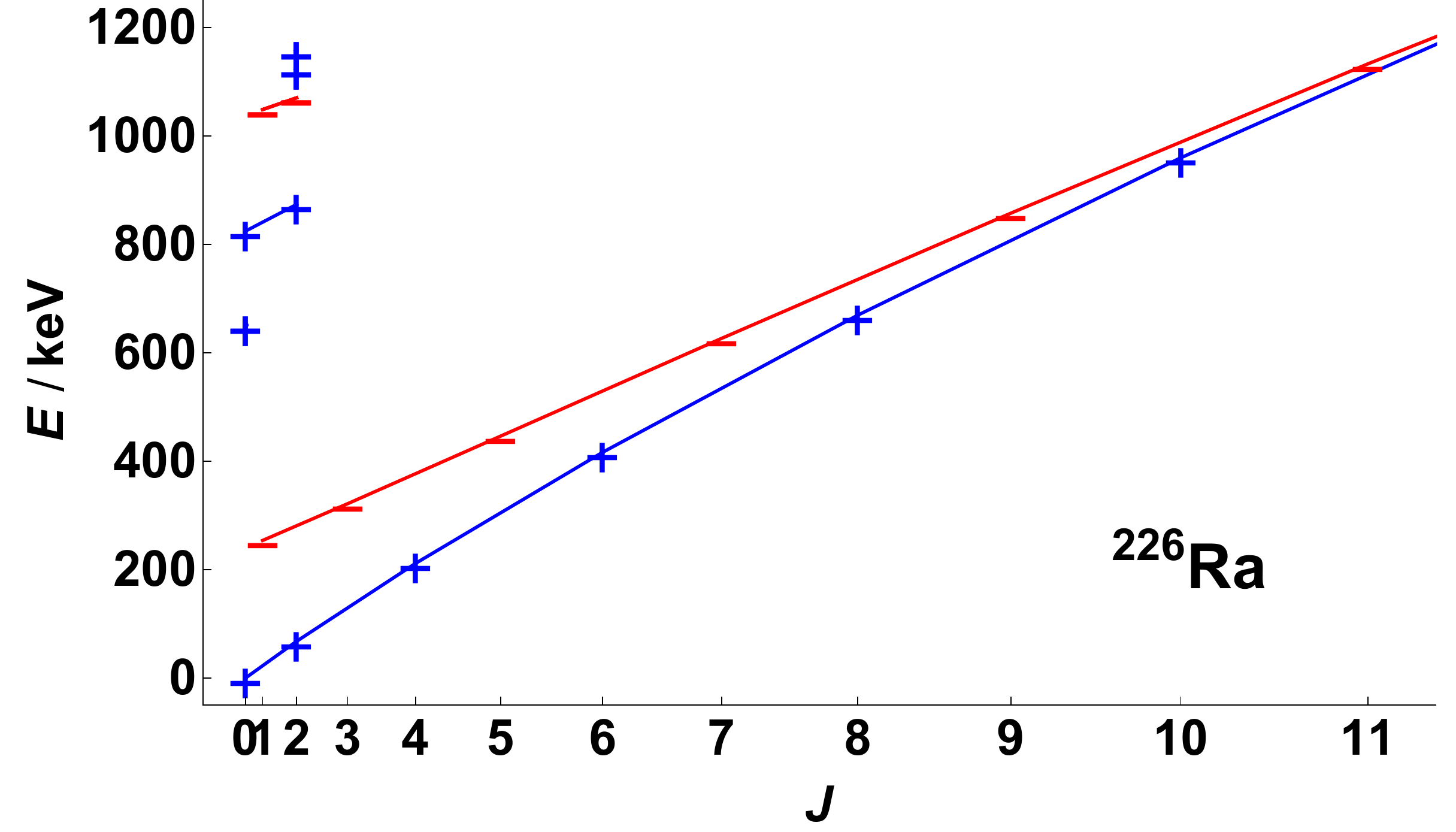}
\caption{\label{fig:226Ra}
Energy levels of $^{226}$Ra as function of $J(J+1)$ for experimentally identified states. Positive and negative parities are indicated by {\color{blue} \bf +} and {\color{red} \bf -- }, respectively. Straight lines indicate rotational bands with $E=E_0+J(J+1)/2{\Theta}$.
}
\end{figure}

$^{227}$Ac (Z=89, N=136) has one more proton than the even-even nucleus $^{226}$Ra. 
In Fig.~\ref{fig:226Ra} one sees a text book example of an octupole rotor spectrum. 
\begin{figure}[h]
\includegraphics[width=0.4\textwidth]{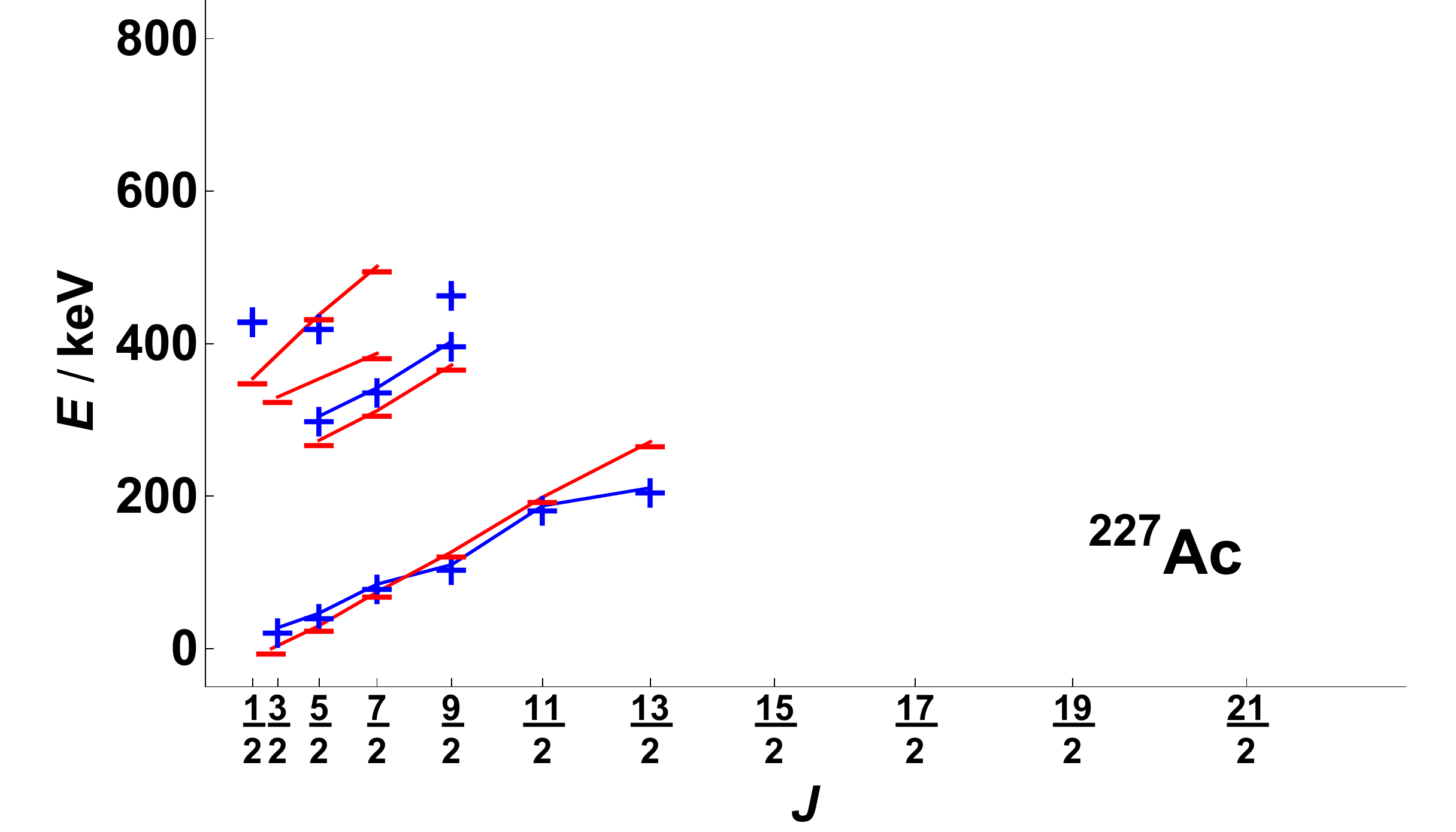}
\caption{\label{fig:227Ac}
Energy levels of $^{227}$Ac as function $J(J+1)$ for experimentally observed states. Positive and negative parities are indicated by {\color{blue} \bf +} and {\color{red} \bf -- }, respectively. Straight lines indicate rotational bands with $E=E_0+J(J+1)/2{\Theta}$. }
\end{figure}

A positive and negative parity band close in energy and even merging for $J \ge 10$.
For small $J$ the collective kinetic energy payed for the collective wave function being antisymmetric in $\beta_3$ amounts only to about 250~keV. As the nucleus stretches with  larger $J$ or larger rotational frequency $\omega$ this collective kinetic energy disappears, indicating that 
$\ket{Q(\omega)}$ and $\op{\Pi}\ket{Q(\omega)}$ have no overlap anymore.
Experimentally this octupole rotor has been identified up to $J = 18$. 
At low energies there seem to be no other rotational bands in the vicinity that obscure the picture.
Adding a proton, which leads to $^{227}$Ac, does not change this octupole signature (Fig.~\ref{fig:227Ac}). Unfortunately there are not so many levels identified experimentally, but close to the ground state one sees for all $J$ close lying parity doublets. 
This and the observation that the energy splitting between the ground state and its parity partner is only 27.4~keV makes $^{227}$Ac the most promising case out of the four presented here.

\section{Conclusion}

 T,P-violating effects (EDM) in  $^{153}$Eu, $^{235}$U, $^{237}$Np and $^{227}$Ac atoms and ions and in molecules containing these atoms are 2-3 orders of magnitude larger than such effects in the 
Hg  atom, 3-4 orders larger in comparison with the Xe atom and  2-3 orders larger in comparison with the TlF molecule where the measurements have been performed. The advantage of  $^{153}$Eu, $^{235}$U,  $^{237}$Np  and $^{227}$Ac in comparison with $^{225}$Ra  (where the magnitude of the atomic EDM  is comparable and the experiment is running) is their stability and availability.  This allows one to do experiments not only with atoms but also with ions and molecules. 
Current accuracy of molecular measurements of the T,P-violating effects should be sufficient for a significant improvement of the limits on the parameters of theories of CP-violation.

Measurements of the effects produced by the Schiff moment may be used to search for the axion dark matter. The  axion dark matter produces oscillating nuclear  EDM and oscillating  Schiff moments which are enhanced  by the same octupole mechanism.

\section{Acknowledgements}. 

   This work is supported by the Australian Research Council grant No. DP150101405 and Gutenberg Fellowship. We are grateful to M. Kozlov,  L. McKemmish, N. Hutzler, A. Palffy, Jun Ye, D. DeMiIle,  N. Minkov, A. Afanasiev, P. Ring, D. O'Donnel, M. Scheck, J. Dobaczewski,  W. Nazarewicz and the TACTICA collaboration for useful discussions.

\end{document}